\documentclass[letterpaper,10pt,conference]{ieeeconf} % IEEE format two column

\IEEEoverridecommandlockouts
\usepackage{graphicx}

\usepackage{amsmath}
\usepackage{amssymb}
\usepackage{amsfonts}
\usepackage{mathtools}
\usepackage{MnSymbol}
\usepackage{subcaption}
\usepackage{amsthm}
\usepackage{algpseudocode}
\usepackage[usenames,dvipsnames,svgnames,table]{xcolor}
\usepackage{float}
\usepackage{caption}
\usepackage{cite}
\floatstyle{plaintop}
\restylefloat{table}
\usepackage[vlined,ruled,linesnumbered]{algorithm2e}
\newtheorem{definition}{Definition}

\newtheorem{problem}{Problem}

\usepackage{multirow}

\newcommand{\levent}{\lozenge}

\newcommand{\luntil}{\mathcal{U}}
\newcommand{\lnext}{\bigcirc}

\newcommand{\asgn}{\leftarrow}

\newcommand{\TS}{\mathcal{T}}
\newcommand{\FA}{\mathcal{A}}

\title{\LARGE \bf
Distributed Fair Assignment and Rebalancing for Mobility-on-Demand Systems via an Auction-based Method
}

\author{Kaier Liang and Cristian-Ioan Vasile
\thanks{Kaier Liang and Cristian-Ioan Vasile  are with the Mechanical Engineering and Mechanics Department at Lehigh University, PA, USA: {\tt\small \{kal221, cvasile\}@lehigh.edu}}        
}     

\renewcommand{\baselinestretch}{0.97}

\begin{document}

\maketitle
\thispagestyle{empty}
\pagestyle{empty}

%%%%%%%%%%%%%%%%%%%%%%%%%%%%%%%%%%%%%%%%%%%%%%%%%%%%%%%%%%%%%%%%%%%%%%%%%%%%%%%%
\begin{abstract}
In this paper, we consider fair assignment of complex requests for Mobility-On-Demand systems.
We model the transportation requests as temporal logic formulas that must be satisfied by a fleet of vehicles. We require that the assignment of requests to vehicles is performed in a distributed manner based only
on communication between vehicles while ensuring fair allocation.
Our approach to the vehicle-request assignment problem is based on a distributed auction scheme with no centralized bidding that leverages utility history correction of bids to improve fairness.
Complementarily, we propose a rebalancing scheme that employs rerouting vehicles to more rewarding areas to increase the potential future utility and ensure a fairer utility distribution.
We adopt the max-min and deviation of utility as the two criteria for fairness. 
We demonstrate the methods in the mid-Manhattan map with a large number of requests generated in different probability settings. 
We show that we increase the fairness between vehicles based on the fairness criteria without degenerating the servicing quality.
\end{abstract}

%%%%%%%%%%%%%%%%%%%%%%%%%%%%%%%%%%%%%%%%%%%%%%%%%%%%%%%%%%%%%%%%%%%%%%%%%%%%%%%%
\section{Introduction}
Mobility-On-Demand systems have been recognized as a promising solution to reduce travel costs, traffic congestion, and emissions~\cite{teubner2015economics, liyanage2019flexible}.
Passengers can specify their demands and share vehicles with others, and it can greatly improve transportation for people and goods.
However, most research in this area has been focused on the passenger's perspective, and less attention has been paid to the problem from the driver's perspective. 
The assignment objectives are usually centered on minimizing the travel cost, which may not be in accord with the driver's preferences~\cite{cao2021optimization, aleksandrov2022fair}. 
Moreover, within the vehicle fleet, due to competition, unfairness may arise due to the uneven distribution of requests, resulting in some vehicles being underutilized.

Furthermore, the vehicle assignment problem is usually done via a centralized method, such as optimization~\cite{foti2017nash, lin2012research, levin2017congestion}. 
However, this method requires drivers to share a lot of information with all other vehicles and adhere to the assignment provided by the centralized solver.
Although fleets belonging to the same company may be willing to follow the centralized assignment, it may not be suitable for situations with numerous competitors or a large number of independent drivers.
As a result, using distributed methods that require vehicles to share limited information with only limited groups can be more favorable~\cite{pandey2019needs, simonetto2019real}. 

Rebalancing policy is also studied to improve efficiency and alleviate congestion problems~\cite{wen2017rebalancing, smith2013rebalancing, spieser2016shared}.
Rebalancing works by moving idle vehicles to another location based on different criteria and purposes, e.g., directly serving other unassigned requests, avoiding congestion, increasing the likelihood of picking up requests, and thus improving performance.
However, rebalancing can be also effective in terms of fairness for the vehicles.
As idle vehicles being mobilized by rebalancing can also receive more utilities in the future.
In this paper, we use the rebalancing approach to improve the fairness for drivers, specifically to balance their collected utilities over a period of time.

Another aspect that has received increasing attention, is the idea of moving from simple pick-up and drop-off requests to more complex demands that do not require customers to plan out trips for their tasks.
This is especially important for unmanned transportation.
Moreover, requests may need to share the same vehicle or use more than one.
To accommodate these two problems, we use Linear Temporal Logic (LTL) to model requests in the vehicle routing problem~\cite{tumova2016least}.
Temporal logics have been successful in specifying and automating the synthesis of control and motion policies for robots~\cite{vasile2013sampling, kress2009temporal, plaku2016motion,9635906} and dynamical systems~\cite{ding2014optimal, 9993288, wolff2014optimization}.
In this work, we leverage automata-based techniques~\cite{belta2017formal} to compute small routing problems with LTL requests, and employ a distributed auction algorithm to assign the requests to vehicles.
% And by constructing product automata to check the ride-sharing and shortest path with the constraints satisfied.

% , the tasks for requests mainly consist of simple pickup and drop-off requests. However, the requests can be more complex than two nodes transportation, especially for unmanned transportation, and also different requests can potentially share the same vehicles. To accommodate these two problems, we can use linear temporal logic (LTL) to model the vehicle routing problem\cite{tumova2016least}. And by constructing product automata to check the ride-sharing and shortest path with the constraints satisfied.

The contributions of this work are the following:
1) We define a distributed auction assignment algorithm with temporal logic demands where all computation is performed based on inter-vehicle communication and no vehicle has a special role (e.g., centralized bidding),
2) We propose a rebalancing scheme to move idle vehicles to more rewarding locations that takes into account fair distribution of vehicles' cumulated utility,
3) We show via case studies in a large environment in mid-Manhattan with a large fleet of vehicles and a number of requests that our distributed assignment method does not degenerate the performance of the Mobility-on-Demand system compared to a centralized approach. Moreover, our algorithms significantly reduce the deviation of utility and increase the minimum utility, which leads to fairer distribution for vehicles.

% compare the results from the centralized method to show that the distributed method does not degenerate the performance of the Mobility-on-Demand system, and that it significantly reduce the deviation utility and increase the min utility to allow a more fair distribution for vehicles.
\vspace{-0.25cm}
\section{Preliminaries}
\vspace{-0.1cm}
In this section, we introduce the notation used in the paper and review concepts in formal language and automata theory.

We denote the set of real and integer numbers as $\mathbb{R}$ and $\mathbb{Z}$, respectively.
The real and integer numbers greater than $a$ are denoted by $\mathbb{R}_{> a}$ and $\mathbb{Z}_{> a}$.
Similarly, we have $\mathbb{R}_{\geq a}$ and $\mathbb{Z}_{\geq a}$ for real and integer numbers greater or equal to $a$.
% We denote the 
For a finite set $S$, we denote its cardinality and the power set as $|\mathrm{S}|$ and $2^S$. %, respectively.

\begin{definition}[Finite Automaton]
\label{def:dfa}
A deterministic finite state automaton (DFA) is a tuple $\FA=\left(Q_\FA, q_{init}^\FA, 2^\Pi, \delta_\FA, F_\FA\right)$, where $Q_\FA$ is a finite set of states; $q_{init}^\FA \in Q$ is the initial state; $2^\Pi$ is the input alphabet; $\delta_\FA : Q_\FA \times 2^\Pi \to Q_\FA$ is a transition function; $F_\FA \subseteq Q_\FA$ is the set of accepting states. 
\end{definition}

An input word $\boldsymbol{\sigma}=\sigma_{0} \sigma_{1} \ldots \sigma_{n}$ over alphabet $2^\Pi$
generates the \emph{trajectory} of the DFA
$\mathbf{q}=q_{0} q_{1} \ldots q_{n}$
with $q_{init} = q_0$ and $q_{k+1} = \delta_\FA(q_k, \sigma_k)$, for all $k \in \{0,\ldots, n-1\}$.
The trajectory $\mathbf{q}$ is called \emph{accepting} if $q_n \in F_\FA$.
\vspace{-0.1cm}
\begin{definition}[scLTL]
\label{def:scltl}
A co-safe Linear Temporal Logic (scLTL) formula over a set of atomic propositions $\Pi$ is defined recursively as:
{
\setlength{\abovedisplayskip}{2pt}
\setlength{\belowdisplayskip}{2pt}
\begin{equation*}
\phi::=\pi \mid \lnot \pi \mid \phi_1 \lor \phi_2 \mid \phi_1 \land \phi_2 \mid \lnext \phi\mid \phi_1 \luntil \phi_2 \mid \levent \phi,
\end{equation*}}
\vspace{-0.1cm}
where $\phi_1, \phi_2$ are scLTL formulae,
$\pi \in \Pi$ is an atomic proposition,
$\lnot$ (negation), $\land$ (disjunction), and $\lor$ (conjunction) are Boolean operators, and $\luntil$ (until), $\lnext$ (next), and $\levent$ (eventually) are temporal operators.
\end{definition}
\vspace{-0.1cm}

The semantics of scLTL formulae are defined over infinite words with symbols from $2^{\Pi}$.
Intuitively, $\lnext \phi$ holds if $\phi$ is true at the next position in the word; $\phi_1 \luntil \phi_2$ expresses that $\phi_1$ is true until $\phi_2$ becomes true; and $\levent \phi$ expresses that $\phi$ becomes true at some future position in the word.
The formal definition of the semantics can be found in~\cite{baier2008modelchecking}.
Given a word $\boldsymbol{\sigma}$ over the alphabet $2^\Pi$ that satisfies the scLTL formula $\phi$,
we denote the satisfaction as $\boldsymbol{\sigma} \models \phi$.
A finite word $\boldsymbol{\sigma}$ satisfies scLTL formula $\phi$ if for all infinite $\boldsymbol{\sigma}'$ the concatenated (infinite) word $\boldsymbol{\sigma}\boldsymbol{\sigma}' \models \phi$.
The finite word $\boldsymbol{\sigma}$ is \emph{minimal} if none of its prefixes satisfies $\phi$.

scLTL formulae can be translated to DFAs using off-the-shelf tools such as scheck~\cite{Latvala03} and spot~\cite{Duret.13.atva}.
\vspace{-0.1cm}
\begin{definition}[Weighted Transition System]
\label{def:wst}
A weighted transition system (WTS) is a tuple $\TS=\left(S, s_{\text {init }}, D, W, \Pi, L\right)$, where $S$ is a finite set of states, $s_{init} \in S$ is the initial state, $D \subseteq S \times S$ is a transition relation, $W: D \to \mathbb{R}_{\geq 0}$ is a weight function, $\Pi$ is a set of atomic propositions and $L: D \to 2^{\Pi}$ is a labeling function.
\end{definition}
\vspace{-0.2cm}
% \cristi{Notation $D$ might not be the best for the transition relation. However, leave it for now as is.}

The transition from the current state $s$ at time $t$ to the next state $s'$ is reached at time $t' = t + W((s, s'))$ if $(s, s') \in D$.
A \emph{trajectory} of $\TS$ is a finite sequence
$\mathbf{s} = s_0 s_1 \ldots s_n$, such that $s_0 = s_{init}$, and $(s_k, s_{k+1}) \in D$ for all $k \in \{0,\ldots, n-1\}$.
The length of the trajectory $\mathbf{s}$ is $n$,
and its total duration is $W(\mathbf{s}) = \sum_{i=0}^{n - 1} W((s_i, s_{i+1}))$.
The \emph{output trajectory} induced by $\mathbf{s}$
is $\mathbf{o} = L(s_0) L(s_1) \ldots L(s_n)$.
A finite trajectory $\mathbf{s}$ satisfies a scLTL formula $\phi$, denoted $\mathbf{s} \models \phi$, if the induced output trajectory $\mathbf{o} = L(\mathbf{s})$ satisfies $\phi$.
\vspace{-0.2cm}
\section{Problem Formulation}
\vspace{-0.05cm}
In this section, we formulate the fair mobility-on-demand problem with requests expressed as scLTL specifications.
The objective is to sequentially generate assignments for incoming scLTL requests to a fleet of vehicles, with the goal of minimizing the total travel time and ensuring fairness among the fleet of vehicles.
\vspace{-0.2cm}
\subsection{Vehicle, Environment, and Request Models}
\label{sec:models}

The fleet of vehicles $\mathcal{V} = \{v_1, v_2, \ldots, v_p\}$ is deployed
in a road network with intersections $S$ and roads $D \subseteq S \times S$.
The transition $(s, s') \in D$ represents a road from intersection $s$ to $s'$.
Each vehicle $v \in \mathcal{V}$ is initially located at $s_{v,init} \in S$.
All vehicles' motion evolves in discrete time $t\in \mathbb{Z}_{\geq 0}$ synchronized via a global clock.
The traversal duration of road $(s, s')$ is $W((s, s')) \in \mathbb{Z}_{>0}$.

Vehicles are tasked with satisfying a finite set of request $\mathcal{R} = \{r_1, r_2, \ldots, r_m\}$
that arrive sequentially over the horizon time $H \in \mathbb{Z}_{>0}$.
A request $r \in \mathcal{R}$ is defined as a tuple
$r = (\pi_{pick, r}, \phi_r, t_{req, r}, \rho_r, \Omega_{\max, r}, \Delta_{\max, r})$, where
\begin{itemize}
    \item $\pi_{pick, r}$ is a proposition marking the pick-up location;
    \item $\phi_r$ is the scLTL formula specifying the request;
    % tasks translated into a deterministic non-blocking automaton $\mathcal{A}_i$.
    \item $t_{req, r} \in \{0, \ldots, H\}$ is the request's arrival time;
    \item $\rho_r \in \mathbb{Z}_{>0}$ is the number of required seats;
    \item $\Omega_{\max, r} \in \mathbb{Z}_{>0}$ is the maximum waiting time, i.e., the latest accepted pick-up time is $t_{req, r} + \Omega_{\max, r}$;
    \item $\Delta_{max, r} \in \mathbb{Z}_{>0}$ is the maximum allowed delay.
\end{itemize}

The maximum transportation capacity of vehicle $v \in \mathcal{V}$ is $Cap_v \in \mathbb{Z}_{>0}$,
while the \emph{available capacity} at time $t$ is $c_v(t) \in \{0, \ldots, Cap_v\}$.
Vehicle $v$ is \emph{available} at time $t$ if $c_v(t) > 0$,
it is \emph{occupied} if $c_v(t) = 0$,
and \emph{vacant} if $c_v(t) = Cap_v$.
The sets of available and vacant vehicles at time $t$ are $\mathcal{V}^a_t$
and $\mathcal{V}^{vac}_t$, respectively.

% Vehicles have limited transportation capacities.
% We denote by $Cap_v \in \mathbb{Z}_{>0}$ and $c_v(t) \in \{0, \ldots, Cap_v\}$
% the maximum capacity
% and the \emph{available capacity} at time $t$
% for vehicle $v \in \mathcal{V}$.
% A vehicle $v$ is said to be \emph{available} at time $t$ if $c_v(t) > 0$,
% otherwise it is \emph{occupied}, i.e., $c_v(t) = 0$.
% The set of available vehicles at time $t$ is denoted by $\mathcal{V}^a_t$.

% A group of vehicles $V \subseteq \mathcal{V}$ completes a request $r \in \mathcal{R}$
% if they pick up $r$
% at the intersection marked with $\pi_{pick, r}$
% such that their overall available capacity
% is greater than $\rho_r$.
% Formally, we have $\mathbf{s}_v \models \tilde{\phi}_r$,
% vehicle $v$ is available at time $t_{pick, r, v}$ for all vehicles $v \in V$,
% and $\sum_{v \in V} c_v(t_{pick, r, v}) \geq \rho_r$,
% where $\tilde{\phi}_r = \levent (\pi_{pick, r} \land \phi_r)$,
% $\mathbf{s}_v$ is the finite trajectory of $v$ and
% $t_{pick, r, v}$ is the pick-up time for 
% $r$ by $v$.
% Note that we do not require all vehicles $V$ to pick up their share of request $r$ simultaneously.

The delay $\Delta_r$ is the difference between the actual and optimal satisfaction duration.
Formally, $\Delta_r = \max_{v\in V} t_{drop,r,v} - t_{req, r} - t^*_r$, 
where $t_{drop,r,v}$ is the drop-off time of request $r$ by vehicle $v$, and $t^*_r$ is the optimal satisfaction time, i.e.,
the minimum duration to fulfill the request if a vehicle picks up the request at $t = t_{req,r}$ and does not share with other requests.
We require that $\Delta_r \leq \Delta_{\max, r}$.

At current time $t \in \mathbb{Z}_{\geq 0}$, a request is \emph{active} if $t_{req} \leq t$ and it has not been picked-up yet;
a request is \emph{in progress} if it has been picked up and not completed.
The sets of active and in progress requests at time $t$
are $\mathcal{R}^a_t$ and $\mathcal{R}^p_t$, respectively.

An assignment $Asg_t: \mathcal{R}^a_t \to \mathcal{V}^a_t$ at time $t = t_{req, r}$
allocates active requests to vehicles.
If the assignment $Asg_t(r) = \emptyset$, then $r$ is \emph{unassigned} at time $t$.
In case this holds for all $t\in \{t_{req, t}, \ldots, t_{req, r} + \Omega_{max,r}\}$, $r$ is unassigned.
Requests that are \emph{in progress} cannot be reassigned, and vehicles need to be \emph{available} before picking up new requests.
Between request arrivals times, i.e., $t\neq t_{req, r}$, assignments do not change.
The travel duration for the vehicle $v$ fulfilling request $r$ starts from
the time $t_{asgmt, r, v}$ when $r$ is assigned to $v$ until $v$ is dropped off at time $t_{drop, r, v}$.
Formally, we have
{
\setlength{\abovedisplayskip}{2pt}
\setlength{\belowdisplayskip}{2pt}
\begin{equation}
\label{eq: travel}
   \sigma_{v}(r) = t_{drop, r, v} - t_{asgmt, r, v},
\end{equation}}
% $\sigma_{v_i}(r_j)$ captures the total serving duration from the time vehicle is assigned to the request to the time the request is dropped off.

Our objective is to minimize the total traveling duration for all requests defined as
{
\setlength{\abovedisplayskip}{2pt}
\setlength{\belowdisplayskip}{2pt}
\begin{equation}
\label{eq:total-cost}
    J = \sum_{r_i \in \mathcal{R}} \sigma_{v_i}(r_i),
\end{equation}
}
where $v_i$ is the vehicle satisfying request $r_i$.

This problem can be solved using centralized methods such as optimization techniques~\cite{alonso2017demand,liang2022fair}.
However, centralized approaches may not be able to handle disruptions well in real-time, e.g.,
vehicles entering and leaving the system and changes in the environment and requests.
These issues are compounded by the need to collect information into a central node for decision making
which may lead to delays.
Moreover, for vehicle-request problems, each vehicle usually makes individual choices,
and vehicles may not be willing to disclose the information to others.
Therefore, in this paper, we seek distributed assignment methods that avoid the need for centralized data collection.

We assume that all agents can communicate with each other, e.g., via broadcasting to the entire fleet or a subgroup of vehicles.
In this paper, a \emph{distributed assignment} $Asg_t$ at time $t$ is defined as a assignment function
computed by each vehicle based on messages exchanged with other vehicles, and no vehicle takes
a special role in decision-making and coordination.

\vspace{-0.1cm}
\begin{problem}[Distributed Assignment]
\label{pb:assignment}
Given the set of vehicles $\mathcal{V}$ deployed in environment $\mathcal{T}$,
and the set of requests $\mathcal{R} = \{r_1, \ldots, r_m\}$ arriving sequentially over time horizon $H$,
compute distributed assignments $Asg_t$ at each sample time $t\in \{0, \ldots, H\}$ 
and routes $\mathbf{s}_v$ for all vehicles $v\in \mathcal{V}$ %via distributed assignment
such that the total servicing time $J$ is minimized.
% where vehicles only share limited information without a centralized data collector.
\end{problem}
\vspace{-0.3cm}
\subsection{Fairness}
For vehicle assignment problems, the serving rate or customer satisfaction is a crucial factor.
However, it is equally important to consider drivers' viewpoints in terms of the fairness of allocating requests.
The utility for a vehicle $v$ for a given time period $h$ is the sum of the onboard passengers:
{\setlength{\abovedisplayskip}{1pt}
\setlength{\belowdisplayskip}{1pt}
\begin{equation}
\label{eq: utility}
    U_{v} = \sum_{t=0}^h (Cap_{v} - c_{v}(t)).
\end{equation}%
}
%
%where $h$ is the serving time for a vehicle $v_i$ to perform the serving request tasks.
Vehicles' utilities may vary greatly over the service horizon $H$.
Thus, it is important to ensure fair assignment of requests while maintaining good
overall performance of the fleet in terms of the total travel time for requests satisfaction $J$.
% It is expected that the utility of all vehicles received can be very different. Therefore, how to make the assignment in a proper fair way while not degenerating the serving quality is very important.

There are different criteria to quantify fairness, such as envy-free fairness, max-min fairness, and proportionality fairness~\cite {brams1996fair}.
In this paper, we use the \emph{max-min utility} and \emph{deviation of utility} as the two quantities
to measure the fairness of the vehicles.

The max-min fairness criterion emphasizes the maximization of the least utility that a vehicle obtains,
i.e., it captures the lower bound or the worst case of utility.
This criterion is widely used in many applications~\cite{young1995equity}.
The deviation of the utility fairness criterion, on the other hand, captures the utility distribution from the perspective of the entire group, as it directly reflects the utility spread among all vehicles.

In the vehicle assignment scenario, multiple factors can contribute to significantly uneven utility results.
Vehicles' location in the road network impacts their chances of picking up requests due to
spatial and temporal variations of requests' arrival.
Secondly, differences in utility between requests and their limited number can lead to
some vehicles servicing high utility requests while others are assigned lower utility ones or not at all.
This may happen even in the case of a uniform probability distribution of requests over space and time.

% One is due to the different locations of vehicles.
% Since the locations of the road map usually come with a different probability for request generation, then it is inevitable that some locations inherently become ideal locations over others in terms of picking up more requests.
% Secondly, even for a road map with a uniformly distributed probability for request generation, we still cannot expect a uniform or close to the uniform utility distribution for vehicles as the utilities provided by the requests themselves are not even, and the number of requests is limited then some vehicles may be assigned with a high utility request while some may be assigned with the request comes with less utility or even not assigned at all.

The first case, due to spatial and temporal variation, rebalancing strategies can be used to mitigate the effects of request arrival differences over the road network.
Rebalancing works by moving idle vehicles to another location to increase their chances of being assigned requests. 
\vspace{-0.2cm}
\begin{problem}[Fair Rebalancing]
\label{pb:fair-rebalancing}
Given the set of vacant vehicles $\mathcal{V}^{vac}$ deployed in environment $\mathcal{T}$,
compute the rebalancing scheme such that the chances of idle vehicles picking up requests in the future increase.
% the idle vehicles can increase the chances to pick up a request in the future.
\end{problem}

For the second case, due to requests' utility differences,
we impose that assignments are distributed in a fair way
in the sense of max-min and deviation of utility criteria.
% we can adjust the distributed assignment scheme to change the way of the assignment generation thus in return reflecting as the utility distribution in a fairer way.
\vspace{-0.2cm}
\begin{problem}[Distributed Fair Assignment]
\label{pb:fair-assignment}
Given the set of vehicles $\mathcal{V}$ deployed in environment $\mathcal{T}$,
and the set of requests $\mathcal{R} = \{r_1, \ldots, r_m\}$ arriving sequentially over time horizon $H$,
compute distributed assignments $Asg_t$ at each sample time $t\in \{0, \ldots, H\}$
and routes $\mathbf{s}_v$ for all vehicles $v\in \mathcal{V}$ in
% a distributed way using only local information
such that $J$ is minimized and vehicles' utilities are allocated fairly.
% the assignment is conducted distributively and only requires local information for the vehicles such that the results can be fairer.
\end{problem}

\medskip
\vspace{-0.3cm}
\noindent
{\bf Summary of the approach.}

For a fixed time sample interval, we conduct an auction for each active request to available vehicles. First, we construct product automata between the motion model (road network) of a vehicle and the DFAs corresponding to the requests. The route is then computed via the shortest path method (e.g., Dijkstra algorithm) applied on the product automaton graph and projection onto the motion model. If the maximum waiting and delay time is permissible, we allow the vehicles to generate the bid for the requests. After assigning the requests to the vehicles based on the auction results,  we conduct a rebalancing for each idle vehicle to move vehicles to more ideal locations.

\vspace{-0.2cm}
\section{Solution}
\label{sec:solution}
\vspace{-0.1cm}
\subsection{Fair Auction Based Assignment Scheme}
The auction algorithm is a widely used approach for solving assignment problems in a distributed manner.
The algorithm consists of two phases: the bidding phase and the assignment phase.
During the bidding phase, each agent (in our case, each vehicle) makes a bid for each item (i.e., request).
Then, during the assignment phase, the item is assigned to the agent with the highest bid.
This process is repeated iteratively until there is no change in the assignment.
The auction algorithm is known to be optimal and has a polynomial runtime for assignment problems \cite{bertsekas1988auction}.

We modify the standard algorithm to account for fair allocation in addition to optimizing an objective function.
In our specific setting, the objective is to minimize the total traveling time, as defined by equation~\eqref{eq:total-cost}, with the requests as the items for auction and the vehicles as the bidders.
To consider fairness, we add an intermediate \emph{Weight Correction Phase} between bidding and assignment.
The auction algorithm we use for our vehicle assignment problem is outlined in Alg.~\ref{alg: auction}.
In the algorithm, we use two communication primitives:
(a) broadcasting function $broadcast(msg, V)$ that sends message $msg$ to all vehicles in $V$, and
(b) receive function $recv(v')$ that returns the message sent by agent $v'$.
We assume that no packages are lost, and they are received in the same order they are sent.
Thus, the receive function $recv()$ is used in blocking mode.

To find the minimum of the objective function, the auction algorithm is used in reverse.
We use the travel time with opposite sign to compute the first and second most rewarding requests in lines 5-6  based on the utility value defined in equation~\eqref{eq: utility}.
Specifically, the vehicles prefer requests that induce lower travel times.
During the bidding phase, each available vehicle places a bid for the most desirable request.
This utility value takes into account the constraints of maximum waiting time and the delay time for the request.
The bid amount is calculated in line 7 and is the sum of the request's price, the difference between the first and second most desirable request's utility difference, and a slack constant variable $\epsilon$.
This constant is typically set as $\frac{1}{N}$, where $N$ is the number of bidders.
The price of a request is initialized with the negative of the smallest travel time of any request for the vehicle at line 3.
Agents broadcast their preferred request (line 8) to the fleet,
and construct the \emph{bidding group} $G$ of other agents interested in the same request (line 9).

After the bidding phase, a weight correction phase is added to promote fairness.
The weight correction is computed using equation~\eqref{eq: weight correction}, which adjusts the original travel utility based on the difference between the vehicle utility and the average utility of all vehicles in the same bidding group $G$.
This allows vehicles with low history utility to increase their bids beyond their actual bidding capability, giving them a greater chance of winning the auction.
The weight correction phase aims to balance the auction and prevent vehicles from continuously dominating the auction process.
{
\setlength{\abovedisplayskip}{2pt}
\setlength{\belowdisplayskip}{2pt}
\begin{equation}
\label{eq: weight correction}
   \mathrm{WeightCorrection}(v, U_{v_{avg}, G}) =  \alpha \cdot (U_v - U_{v_{avg}, G}),
\end{equation}}
where $\alpha$ is a constant tuning parameter and $G$ is a bidding group of vehicles.
$U_{v_{avg}, G}$ is the average history utility for all $v \in G$.
We employed the weight correction in our previous integer linear programming (ILP) approach~\cite{liang2022fair}, which requires all vehicles to send their history utility to a central node.
However, since our goal is to have a distributed implementation, we restricted the weight correction to be performed only within the same bidding group.
This means that vehicles that bid on the same request adjust their bids only locally inside the group.
This modification enables us to maintain the distributed nature of our approach.
The communication between agents in the bidding group $G$ is captured in lines 10-11 of Alg.~\ref{alg: auction}.
Vehicles within $G$ exchange their utility histories computed using equation~\ref{eq: utility} to compute the mean utility value of the group (line 12).

Finally, during the assignment phase, the request is allocated to the vehicle that offers the highest bid (lines 14-17), and the auction is executed iteratively.
In the subsequent rounds, other vehicles can increase their bids until the highest bid and bidder remain the same.
Note that the price of the request is also updated at the end of each round at line 18.

\setlength{\textfloatsep}{0pt}
\begin{algorithm}[htb]
\caption{Fair Auction Algorithm}
\label{alg: auction}
\small
% \footnotesize
\KwIn{$\TS$ -- the road map, $\mathcal{V}^a$ -- available vehicles, $\mathcal{R}$ - active requests }
\KwOut{$Asg_t$ -- Assignment Function}
\DontPrintSemicolon
\BlankLine

\ForEach{$v \in \mathcal{V}^a$}{
    \tcp*[l]{Initialization}
    Compute $\sigma_{v}(r)$ for each $r\in \mathcal{R}$ using $\mathcal{P}_r = \TS \times \FA_r$\;
    $p_{j} \asgn -\min_{r \in \mathcal{R}} \sigma_{v}(r)$ \tcp*[l]{Set initial price}

    \While{$Asg_t$ changed}{
    
    \tcp*[l]{I. Bidding Phase}
    % \tcp*[l]{Find the most rewarding Request j}
    $U_{v,j} \asgn \max_{r_j \in \mathcal{R}}(-\sigma_{v}(r_j))$ \tcp*[l]{Find the most rewarding Request j}
    % \tcp*[l]{Find the second most rewarding Request k}
    $U_{v,k} \asgn \max_{r_k \in \mathcal{R}\setminus \{r_j\}}(-\sigma_{v}(r_k))$ \tcp*[l]{Find the second most rewarding Request k}
    % \tcp*[l]{Compute initial bid}
    $B_{v, j} \asgn p_{j} + U_{v,j} - U_{v,k} + \epsilon$ \tcp*[l]{Initial bid}
    broadcast($r_j$, $\mathcal{V}^a$)\;
    \tcp*[l] {Group of vehicles bids for $r_j$}
    $G \leftarrow \{v\} \cup \{v' \in \mathcal{V}^a \setminus \{v\} \mid recv(v') = r_j \}$\;
    \BlankLine

    \tcp*[l]{II. Weight Correction Phase}
    broadcast($U_v$, $G$) \tcp*[l] {Broadcast to $G$}
    $U^G_{v'} \asgn recv(v'), \forall v'\in G\setminus\{v\}$ \; %\tcp*[l] {Collect utilities}
    \tcp*[l] {Compute mean utility history for $G$}
    $U_{v_{avg}, G} \asgn \frac{1}{|G|}\sum_{v' \in G} U^G_{v'}$\;
    \tcp*[l] {Update bid}
    $B_{v, j} \asgn B_{v, j} + \mathrm{WeightCorrection}(v, U_{v_{avg}, G})$\;
    \BlankLine

    \tcp*[l]{III. Assignment Phase}
    broadcast($B_{v, j}$, $G$) \tcp*[l]{Broadcast bid to $G$}
    $B_{v', j} \asgn recv(v'), \forall v' \in G \setminus \{v\}$\;
    $v^* \asgn \arg\max_{v' \in G}{B_{v', j}}$ \;
    $Asg_t(r_j) \asgn v^*$ \tcp*[l]{Assign request to the largest bidder}
    $p_j \asgn B_{v^*, j}$ \tcp*[l]{update price}
    
    }
}
\Return $Asg_t$
\end{algorithm}
It is important to note that even though the auction algorithm restricts a vehicle to bid for only one request at each round, we can still enable vehicle sharing by allowing vehicles with $c_v > 0$ to participate in the next auction, as long as the total capacity does not exceed the maximum $Cap_v$~\cite{simonetto2019real}.
% \vspace{-0.1cm}
\subsection{Automata-based Route Planning}
\vspace{-0.1cm}
To conduct an auction in the bidding phase, we need to determine which vehicles are eligible to bid for which requests and what the utility (essentially the route) is for each request.
We obtain this information through the construction of product automata.

The requests are represented as scLTL formula and vehicles are represented as a WTS.
Formally, we have the $\TS_v=\left(S, s_{\text {init }}, D, W, \Pi, L\right)$
that captures vehicle $v$'s motion in the environment.
The set of propositions $\Pi$ includes the active requests' pick-up propositions $\pi_{pick, r}$.

For every available vehicle $v$ and active request $r$, we construct a weighted product automaton $\mathcal{P}_{rv} = \mathcal{T}_v \otimes \mathcal{A}_r$. $\otimes$ is a product operation.
$\mathcal{T}_v$ is the transition system for the vehicle $v$ with initial position $s_{init}$ set as the vehicle' current position. 
If the vehicle already has an onboard passenger $r'$, we construct the weighted product automaton $\mathcal{P}_{v,r,r'} = \mathcal{T}_{v} \otimes \mathcal{A}_r \otimes \mathcal{A}_{r'}$ to validate if the $r$ can be served together without violating the constraints for $r$ and $r'$.
After obtaining the product automata, we use graph search methods such as Dijkstra’s algorithm to compute an admissible path\cite{sniedovich2006dijkstra}.

The formal definition of the product automaton is the following.
\vspace{-0.2cm}
\begin{definition}[Weighted product automaton at time $t$]
The weighted product automaton $\mathcal{P} = \mathcal{T} \otimes \mathcal{A}_{1} \otimes \ldots \otimes \mathcal{A}_{m}$ of vehicle $v$ at time $t$ is a tuple $\left(Q_{\mathcal{P}}, Q_{\text {init}, \mathcal{P}}, \delta_{\mathcal{P}}, F_{\mathcal{P}}, W_{\mathcal{P}}\right)$, where
\begin{itemize}
    \item $Q_{\mathcal{P}} \subseteq S \times Q_{\mathcal{A}_1} \times \ldots Q_{\mathcal{A}_m}$; %$$ \{s, q_1, \cdots, q_m\}$;
    % \item $Q_{init} = \{s_j, \pi_{pick,1}, \cdots, \pi_{pick,m} \}$, where $s_j$ is the current node of the $v_k$ in the map; \\%, i.e., $s_0 = s_{init}$; \\
    \item $Q_{init} = \{s_j, q_{1, k}, \cdots, q_{m, k} \}$, where $s_j$ is the current state of vehicle $v$ in the map; \\
    $q_{i, k}=
    \left\{\begin{array}{l}
        \delta_{i}\left(\pi_{pick, i}, L\left(s_{j}\right)\right)\text{ if } t_{pick, r_i} = t\\
        \delta_{i}\left(q_{i, k-1}, L\left(s_{j}\right)\right)\text{ if } t_{pick, r_i} < t\\
        q_{init}^{\FA_i} \text{ else,}
    \end{array}\right.$,\\
where $t_{pick,r_i}$ is the pick-up time for $r_i$, $k$ is the current (event) step associated with time $t$,
and $q_{i, k-1}$ are the states of the request at the previous step;
\item $\delta_{\mathcal{P}} \subseteq Q_{\mathcal{P}} \times Q_{\mathcal{P}}$ is a transition function: \\
$\left(\left(s, q_{1}, \ldots, q_{m}\right),\left(s^{\prime}, q_{1}^{\prime}, \ldots, q_{m}^{\prime}\right)\right) \in \delta_{\mathcal{P}}$
if and only if $\left(s, s^{\prime}\right) \in R$ and $\left(q_{i}, L\left(s^{\prime}\right), q_{i}^{\prime}\right) \in \delta_{i}$;
\item $F_{\mathcal{P}}=\left\{\left(s, q_{1,k}, \ldots, q_{m,k}\right) \mid q_{i,k} \in F_{i}, \forall i \in \left\{1, \ldots, m\right\} \right\}$;
\item $ W_{\mathcal{P}}$: $\delta_{\mathcal{P}} \rightarrow \mathbb{R}_{+}$ is the weight function given by $W_{\mathcal{P}}(\left(\left(s, q_{1}, \ldots, q_{m}\right),\left(s^{\prime}, q_{1}^{\prime}, \ldots, q_{m}^{\prime}\right)\right)) = W(s, s^\prime)$.
\smallskip
\end{itemize}
A satisfying path $\mathbf{q}$ in $\mathcal{P}$ connects the initial state $Q_{init}$ with a reachable final state $F_\mathcal{P}$.
If such a path exists, we project it onto $\TS$ by taking the first component of each state in the state path $\mathbf{q}$.

% \begin{aligned}
% &F_{\mathcal{P}}=\left\{\left(s, q_{i_{1}}, \ldots, q_{i_{m_{j}}}, \emptyset\right) \mid q_{i} \in F_{i}, \forall i \in\right
% \left.\left\{i_{1}, \ldots, i_{m_{j}}\right\}\right\}
% \end{aligned}

\end{definition}
\begin{figure*}[hbt!]
    \centering
    \includegraphics[width=0.65\linewidth]{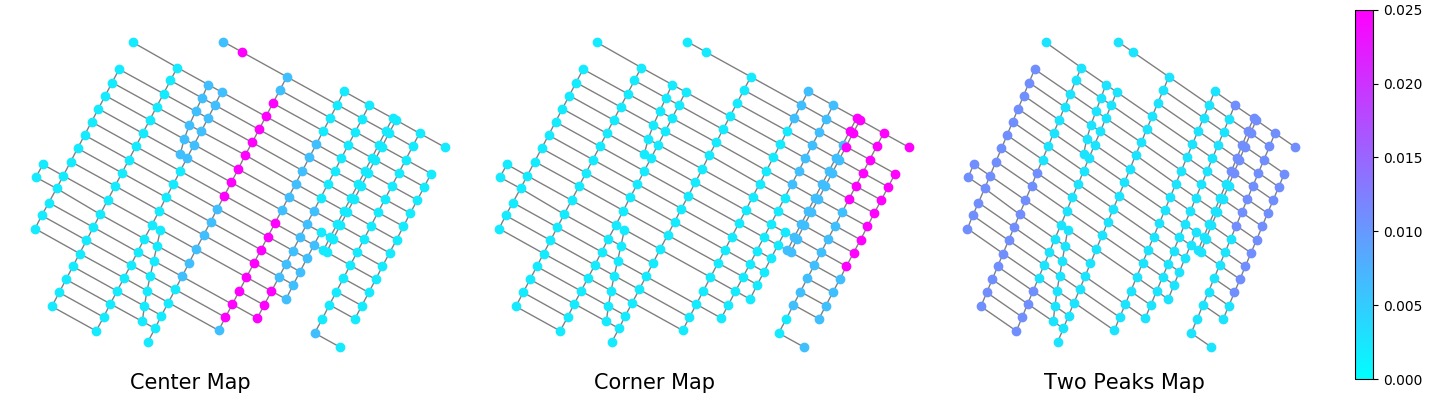}
    \caption{Different settings for the Mid-Manhattan Map: the colors in nodes represent the probability of a new request arrival.}
    \vspace{-0.67cm}
    \label{fig:my_label}
\end{figure*}
\vspace{-0.4cm}
\subsection{Fair Rebalancing}
\vspace{-0.1cm}
In real-life scenarios, the road map for request generation is often non-uniformly distributed. For instance, certain areas like the city center or airport have a higher probability of generating requests than rural areas where requests are infrequent. As a result, due to the maximum waiting time and maximum allowed delay, there can be a significant difference in utility among vehicles, leading to an unfair distribution of utility for the vehicles. To address this problem, we propose a rebalancing scheme that reduces these unfair effects.
For each node $s \in S$ in the road map $\TS$ at time $t$, we calculate the potential utility $P_{t}(s)$ as:
{\setlength{\abovedisplayskip}{2pt}
\setlength{\belowdisplayskip}{2pt}
\begin{equation}
\label{eq:re_utility}
    P_{t}(s) = \frac{Pr_t(s)}{1 + N_{v, t}} * U_{r_{avg}},
\end{equation}}
where $Pr_t(s)$ is the probability of a request arriving at node $s$ at a given time $t$; $N_{v,t}$ is the number of nearby idle vehicles at time $t$ for a fixed distance range, and $U_{r_{avg}}$ is the average utility for requests arriving at $s$ which can be obtained from the history data.
The term $N_{v,t}+1$ is added in the denominator to avoid division by zero.
This formula considers both the probability of a new request arriving and the number of competing vehicles nearby, reflecting the potential utility for a vehicle at location $s$ and time $t$.

We simplify the problem by assuming $Pr_t(s)$ is independent of time and only related to locations.
However, for a large road map and a significant number of vehicles, simply calculating the highest utility $P_{v, t}(s)$ for every vehicle $v$ and rebalancing the vehicle to the corresponding location $s$ can be expensive and inadvisable for several reasons:

\noindent
(1) The highest location can be the same for all vehicles, which can be seen from the independence with respect to a specific vehicle in equation~\eqref{eq:re_utility}.

\noindent
(2) Rebalancing itself will require some cost as it will require idle vehicles to move to another location. Therefore the highest potential location that is far away may be less attractive than a location with a smaller value but close.

To deal with these problems, we use a slack parameter and a distance search window. The rebalancing target location is calculated in Alg.~\ref{alg:find_target}:

\begin{algorithm}
\caption{Fair Rebalancing Algorithm}
\label{alg:find_target}
\KwIn{$\TS$ -- the road map, $\mathcal{V}^{vac}$ -- available vehicles, $k_{w}$ -- Distance Search Window}
\KwOut{$Reb_t$ -- Rebalance Function}
\DontPrintSemicolon
\BlankLine
\small
% \tcp*[l]{Sort the near nodes by degrees in the search window }
\ForEach {$v \in \mathcal{V}^{vac}$}{
$\mathcal{G}_{v} \leftarrow Neighbor(s_{v,init}, \TS, k_{w}(v))$\\
% $s_{v, tar} = s_{v, init}$\\
$P_{v, t}(s_{v, tar}) \leftarrow P_{v, t}(s_{v, init})$\\
$s_{v, tar} \leftarrow s_{init}$\\
\ForEach {$N_{deg,i} \in \mathcal{G}_{v}$}{
$P_{v, t}(s_{v, max}) \leftarrow \max_{s_j \in N_{deg,i}} P_{v,t}(s_j)$\\
\lIf{$P_{v, t}(s_{v, max}) \geq P_{v,t}(s_{v,tar})*k_{a}$}
{
    $s_{v, tar} \leftarrow s_{v, max}$\\
    $P_{v, t}(s_{v, tar}) \leftarrow P_{v, t}(s_{v, max})$
  }
}
$Reb_t(v) \leftarrow s_{v, tar}$
}
\Return $Reb_t$
\end{algorithm}
To implement the rebalancing scheme, we first sort all potential rebalancing locations based on their degree or distance from the initial location, as specified in line 2 of Alg.~\ref{alg:find_target}. The distance search window $k_w(v)$ is used for vehicle $v$ to ensure that the rebalancing search is not performed to a location that is too far away preventing making unnecessary searches. Then in line 5, starting from the first-degree nodes, or the nearest nodes, we then find the maximum potential utility node using equation~\eqref{eq:re_utility}. To ensure that vehicles take account of both the distance and utility, we use a constant slack variable $k_a > 1$ in line 7 to increase the perceived cost of rebalancing to outer degree nodes or farther nodes. The vehicle will only choose to rebalance to a high degree node if the potential utility is significantly greater than the current target rebalancing node.

The auction and rebalancing are implemented sequentially. At a given time sample frequency, an auction is conducted to assign available vehicles to every unassigned request. Then the rebalancing is conducted to move idle vehicles to move to better locations. Therefore, vehicles are either in progress to serve requests or in rebalancing to move to another location.

\vspace{-0.15cm}
\section{Simulation}
\vspace{-0.12cm}
In this section, we present the simulation results to demonstrate the performance of distributed fair assignment and the rebalancing scheme. 
\vspace{-0.22cm}
\subsection{Simulation Specifications}
\begin{figure}[hbt!]
    \centering
\includegraphics[width=1\linewidth]{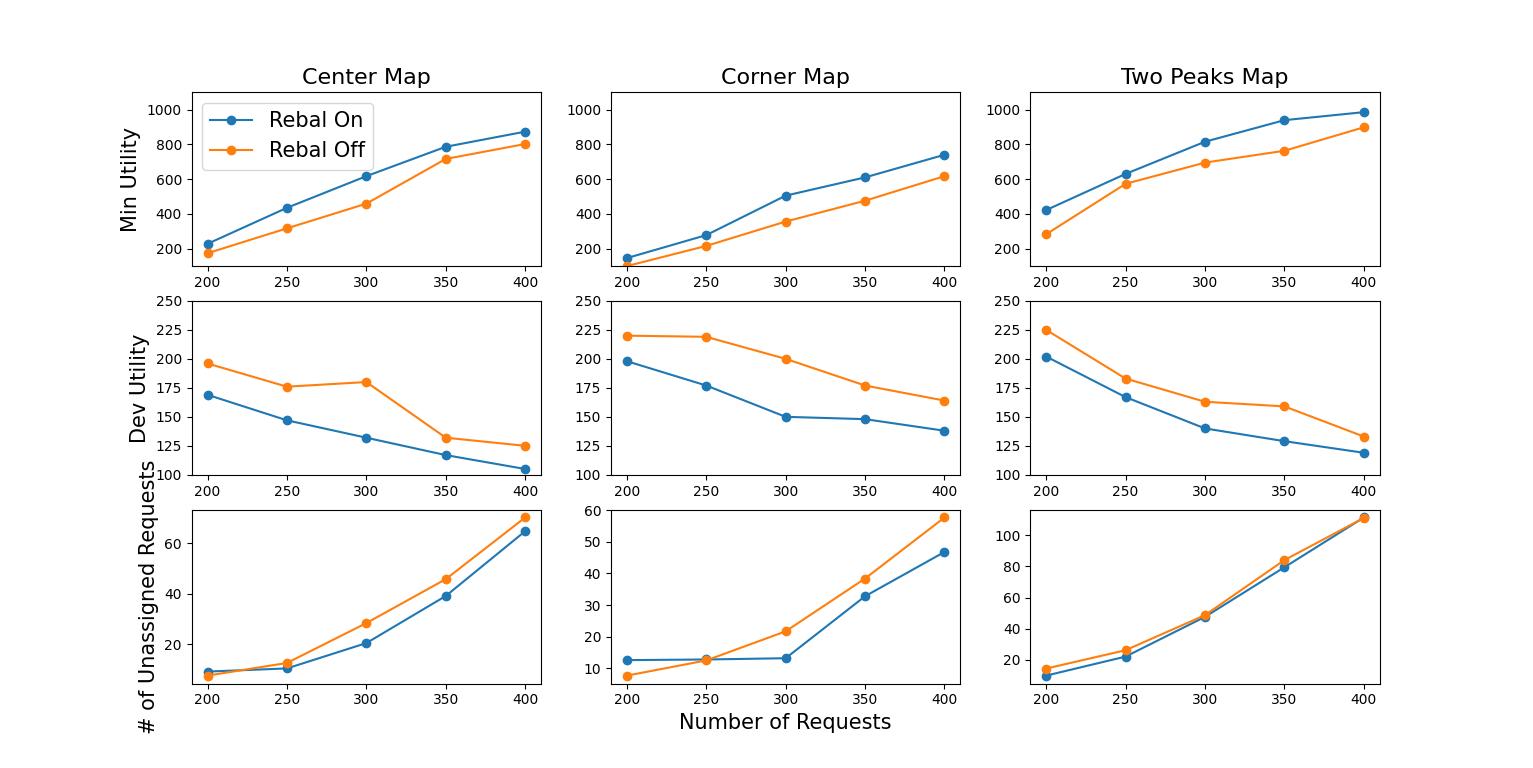}
    \caption{rebalancing performance in different map settings}
    \label{fig: diff maps}
    \vspace{-0.18in}
\end{figure}

\vspace{-0.12cm}
The road map for the simulation is used as the Mid-Manhattan map, which contains 184 nodes, and the weights for every edge are acquired by real travel duration from taxi driving data \cite{alonso2017demand}. We varied the request generation probabilities and the number of requests to evaluate the fairness performance of the system. Three different maps were used for the simulations, namely the center map, corner map, and two peaks map, with request generation probabilities as shown in Fig.~\ref{fig: diff maps}. These maps are characterized by high probability areas where requests are more likely to be generated. So that it can reflect the uneven distribution of requests in real-life scenarios.

The simulation duration is set to 1000 seconds with varying the number of vehicles and requests. The initial positions of all vehicles are generated in a uniform distribution. The scLTL formulas for the requests are generated from the following scLTL pattern stochastically. 
% \vspace{-0.45cm}

% \begin{figure}[hbt!]
%     \centering
%     \subfloat[\label{subfig:min}]{%
%     \includegraphics[width=0.75\linewidth]{figs/min_new.png}}
%     \vspace{-0.01in}
%     \centering
%     \subfloat[\label{subfig:dev}]{%
%        \includegraphics[width=0.75\linewidth]{figs/deviation_new.png}
%        }
%     \vspace{-0.01in}   
%     \centering
%     \subfloat[\label{subfig:avg}]{%
%        \includegraphics[width=0.75\linewidth]{figs/average_new.png}
%        }
%     \caption{Performance comparison for rebalancing and weight correction (center map: 20 vehicles, 300 requests) (a)minimum utility comparison(b)utility deviation comparison(c)average utility comparison}
%     \label{fig: diff pare}
% \end{figure}

\begin{figure}
    \centering
\includegraphics[width=1\linewidth]{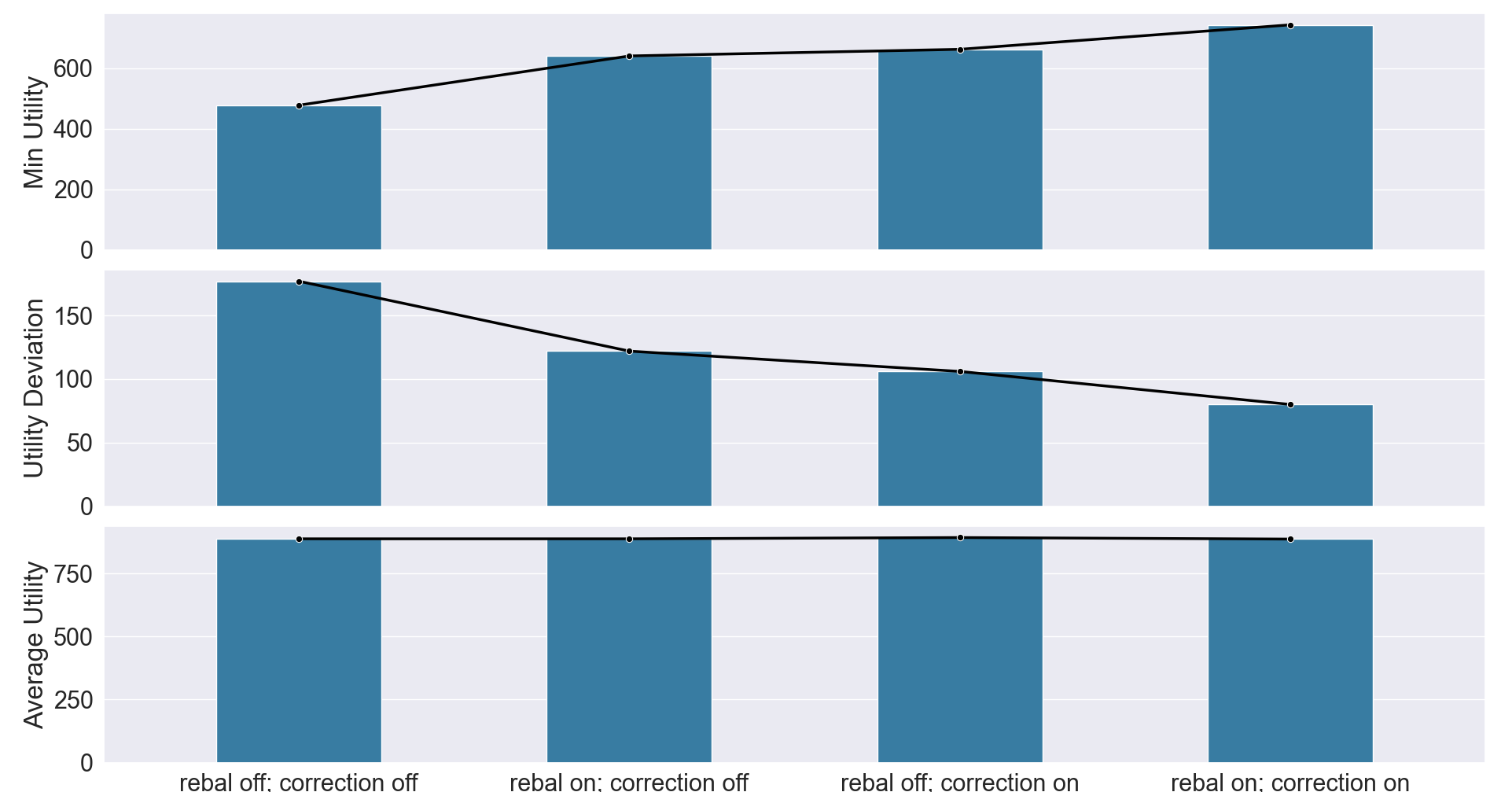}
    \caption{Performance comparison for rebalancing and weight correction (center map: 20 vehicles, 300 requests) from the minimum utility, utility deviation and average utility}
    \label{fig: diff pare}
\end{figure}

\textbf{scLTL pattern:}
\vspace{-2mm}
\begin{equation*}
\begin{gathered}
    \begin{aligned}
    \tilde{\phi}_{1}\left(s_{pick}, s_{1}, s_{2}\right) &= \levent(s_{pick} \wedge \levent\left(s_{1} \wedge \levent\left(s_{2}\right)\right)),\\
    \tilde{\phi}_{2}\left(s_{pick}, s_{1}, s_{2}\right) &= \levent(s_{pick} \wedge\levent\left((s_{1} \vee s_{2}\right) \wedge s_{3})),\\
    \tilde{\phi}_{3}\left(s_{pick}, s_{1}, s_{2}, s_{3}\right) &= \levent(s_{pick} \wedge \levent\left(s_{1} \wedge  (s_{2} \vee s_{3}\right)))
    \end{aligned}
\end{gathered}
\end{equation*}
where $s_{i}$ are locations in the road map. The arrival time is generated according to a uniform Poisson process. The locations are chosen based on the corresponding probability of request generation in the road map. 

Throughout the simulation, we perform the auction and rebalancing every 10 seconds. Additionally, we set the maximum waiting time and delay time to 40 and 100 seconds.

\vspace{-0.2cm}
\subsection{Simulation Results}
\vspace{-0.1cm}
In the simulation results shown in Fig.~\ref{fig: diff maps}, we consider 20 vehicles and a varying number of requests from 200 to 400 to demonstrate the effect of the rebalancing strategy. Each data point in the figure is the average result of 20 simulations. The fairness is compared using the minimum and deviation utility. Fig.~\ref{fig: diff maps} shows we can increase the minimum utility and decrease the deviation utility consistently without degenerating the serving rate in all three map settings.

Fig.~\ref{fig: diff pare} shows the comparison between the planning with and without weight correction and rebalancing settings. 
We can see the improvement of introducing rebalancing or the  weight correction from the two fairness criteria; the settings that adopt the rebalancing or weight correction can increase the minimum utility and decrease the deviation utility. And the setting performs best when it uses the rebalancing and the weight correction together. 

In Fig.~\ref{fig: diff pare}, we also notice that using balancing or weight correction does not affect the average utility. This suggests that although we cannot increase the total utility for the entire system, we can adjust the utility distribution in a fair way by increasing the minimum utility and decreasing the deviation.

\vspace{-0.4cm}
\begin{figure}[hbt!]
    \centering   \includegraphics[width=0.8\linewidth]{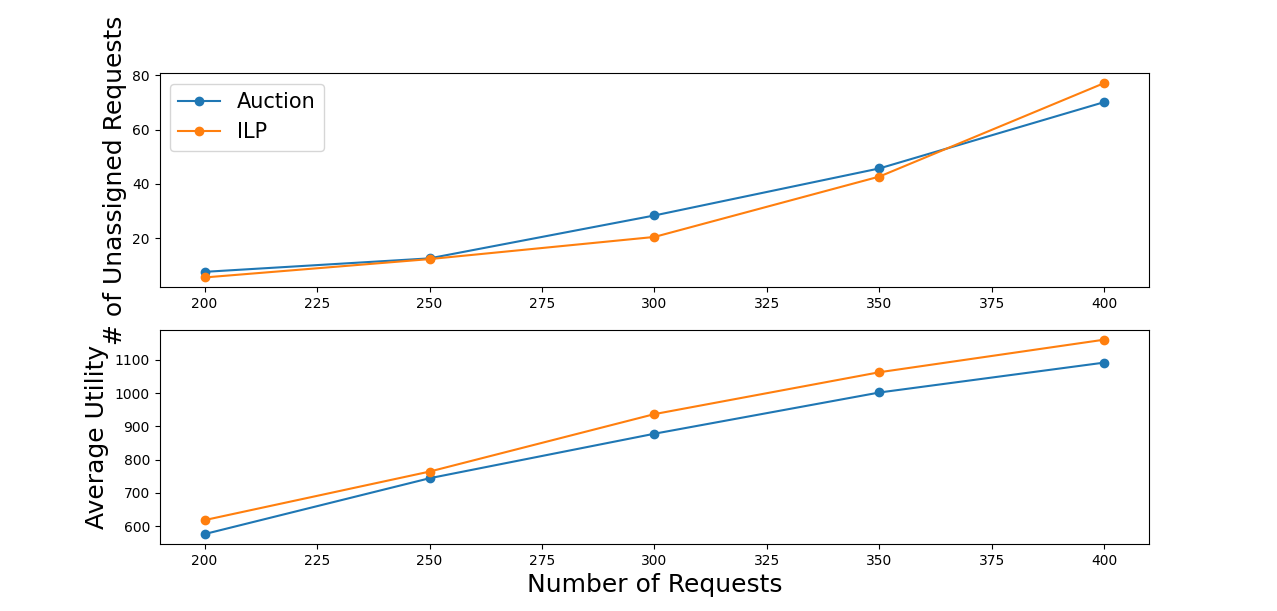}
    \caption{Comparison between ILP and auction without using rebalancing and weight correction}
    \vspace{-0.1cm}
\label{fig: compare}
\end{figure}

Comparison with the centralized approach: here we present the performance comparison between the auction algorithm and centralized algorithm using ILP\cite{liang2022fair}. Although both algorithms can obtain the optimal solution, the algorithms' implementations are different. First, the ILP setting allows more than one request to be assigned together at one step due to the optimization nature, whereas the auction algorithm can only assign one request to one vehicle at one-time. Furthermore, since both methods are run continuously throughout the simulation, it is not possible to obtain the global optimal solution. Therefore, the current optimal solution does not imply the global property, as future events cannot be predicted at the current time step.

For the comparison shown in Fig.~\ref{fig: compare}, we compare the auction and ILP methods for the setting with 20 vehicles and a varying number of requests. While both approaches are to minimize the traveling time, this is not easy to quantify and compare directly. Thus, we evaluate the average utility and the number of unassigned requests. The average utility and the number of unassigned requests capture the running quality from the requests and vehicles' perspectives.
For the comparison, we used both the auction and ILP settings without rebalancing and weight correction.
We see that the two approaches perform very similarly which is expected.

\vspace{-0.2cm}
\section{Conclusions}
\vspace{-0.1cm}
In conclusion, this paper presents a novel approach to the problem of fair assignment and rebalancing in Mobility-On-Demand systems. Our proposed distributed assignment method reduces the need for a central authority for coordination. The introduction of the rebalancing scheme leads to a fairer distribution of requests for vehicles, as demonstrated by an increase in the minimum utility and a decrease in the utility deviation compared to the baseline. By modeling requests using temporal logic formulas, our approach accommodates complex demand patterns. The results of our study demonstrate the efficacy of the proposed method in achieving fairer vehicle assignment in Mobility-On-Demand systems. 

  % This command serves to balance the column lengths
                                  % on the last page of the document manually. It shortens
                                  % the textheight of the last page by a suitable amount.
                                  % This command does not take effect until the next page
                                  % so it should come on the page before the last. Make
                                  % sure that you do not shorten the textheight too much.

%%%%%%%%%%%%%%%%%%%%%%%%%%%%%%%%%%%%%%%%%%%%%%%%%%%%%%%%%%%%%%%%%%%%%%%%%%%%%%%%

%%%%%%%%%%%%%%%%%%%%%%%%%%%%%%%%%%%%%%%%%%%%%%%%%%%%%%%%%%%%%%%%%%%%%%%%%%%%%%%%

\bibliographystyle{ieeetr}
\bibliography{reference.bib}

\end{document}